\documentclass[%
 reprint,
 amsmath,amssymb,
 aps,
]{revtex4-1}

\usepackage{graphicx}
\usepackage{dcolumn}
\usepackage{bm}

\begin{document}

\title{Creating universes with thick walls}
\author{Andrew Ulvestad}
 \affiliation{Physics Department, University of Chicago}
 \email{andrewulvestad@uchicago.edu}
 \author{Andreas Albrecht}
 \email{ajalbrecht@ucdavis.edu}
\affiliation{Physics Department, UC Davis}

\date{\today}

\begin{abstract}
We study the dynamics of a spherically symmetric false vacuum bubble
embedded in a true vacuum region separated by a ``thick wall'', which
is generated by a scalar field in a quartic potential. We study
the ``Farhi-Guth-Guven'' (FGG) quantum tunneling process by constructing numerical
solutions relevant to this process. The ADM mass of the spacetime is calculated, and we show that
there is a lower bound that is a significant fraction of the scalar
field mass.  We argue that the zero mass solutions used to by some to argue
against the physicality of the FGG process are artifacts of the thin
wall approximation used in earlier work. We argue that the zero mass
solutions should not be used to
question the viability of the FGG process. 

\end{abstract}

\pacs{}
\maketitle

\section{Introduction}
While our universe appears to be well described by $\Lambda$CDM
cosmology and slow-roll inflation, much about the pre-inflationary
universe remains speculative. Numerous models rely on quantum
tunneling from some previous state to give an inflating universe that
eventually leads to the universe we observe today (see for example \cite{dse}). 

We consider here the Farhi-Guth-Guven'' (FGG) process which 
was originally studied in the ``thin wall'' limit
\cite{fgg},\cite{fmp}. In this process, a bubble of false vacuum, known as the
seed bubble, is separated by a thin domain wall from a region of true
vacuum. Einstein's equation implies two distinct solutions for the motion of the bubble wall; the first
eventually collapses while the second expands indefinitely
\cite{bgg}. The possibility of tunneling between these two states is
considered. Although FGG consider the case where a seed somehow forms
in Minkowski space, other cases were considered (for example
in \cite{albrechtsorbo}) where the seed forms from Hawking radiation
in de Sitter space. Either way, the seed collapses into a black hole
but hidden behind the black hole horizon is the expanding
solution. The mass of this bubble, $M$, is the $m$ parameter in the
usual Schwarzschild metric, the ``ADM'' mass. 

FGG is known to dominate over Coleman-de Luccia type tunneling\cite{Aguirre:2005nt} and it
has been argued that this process can
produce inflating universes that do not originate from classical
singularities \cite{fgg,fmp,aj}. Despite these features, many calculations
that study tunneling in cosmology (for example in the string theory
``landscape'' \cite{Kachru:2003aw}) ignore FGG, primarily because of various
arguments that this process might not be physical \cite{Freivogel:2005qh,Aguirre:2005nt}. In this
paper we address one of the arguments against the physicality of the
FGG process, one that involves taking the bubble mass $M$ to zero\cite{Aguirre:2005nt}. 
 \begin{figure}[b]
\includegraphics{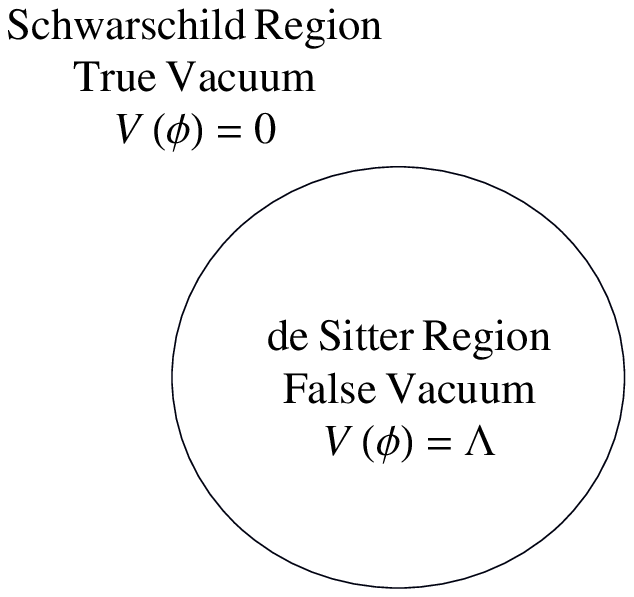}
\caption{\label{fig:diag} A sketch of a bubble solution.}
\end{figure}

The $M\rightarrow0$ limit of the thin wall formula leads to a
prediction of the FGG process that the probability of transitioning
from the seed bubble to the inflating bubble remains finite even as
the mass of the seed bubble is taken to zero. This is the ultimate
free lunch, since it implies our universe was possibly nucleated from
zero matter Minkowski space. However, here we argue that this limit is
an artifact of the thin wall approximation which assumes that the
thickness of the domain wall is small compared to the radius of the
bubble. Indeed, as the radius of the bubble is taken to zero (as it is
in the $M\rightarrow0$ limit), one should expect the thin wall approximation to breakdown. 

In this article we examine bubbles of
false vacuum separated by a ``thick wall'', i.e. scalar
field solutions that interpolate between regions of true and false vacuum.
We construct numerical
solutions for the scalar field coupled to gravity that
are relevant to the FGG process. Probably our most important point is an extremely simple one: For a
fixed potential the types of possible bubbles are limited and the
$M\rightarrow0$ cannot even be taken. So if one has a particular
scalar field potential in mind one is unlikely to encounter the issues
raised in \cite{Aguirre:2005nt} about FGG. 

 In this work we go beyond
this simple point by exploring the parameter space of a general quartic 
scalar field potential (with an overall scale fixed).  We find solutions in such potentials
cannot approach the step-function type solutions for
$\phi$ that are assumed in the thin wall case, even when the potential
is made as ``thin wall'' as possible. Instead, the scalar field
inevitably ``spills over'' and zero mass solutions are unattainable.
By comparison, the thin wall $M\rightarrow0$ limit relies on exact
Schwarzschild space outside the bubble while taking the bubble radius
to zero. While it may be possible to find an exotic potential
  with $M$ arbitrarily small, we show that no quartic potential with a fixed overall
  scale admits such solutions. 

\begin{figure}[b]
\includegraphics{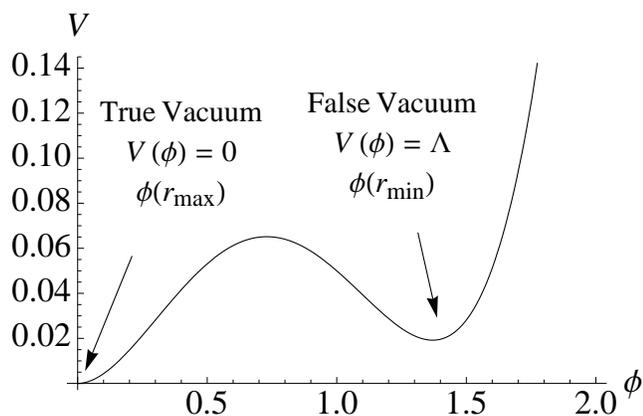}
\caption{\label{fig:potential} A particular $\phi^4$ potential with true and false vacuum regions labeled.}
\end{figure}

\section{The thin wall revisited}

\subsection{The setup}
Imagine embedding a spherically symmetric bubble of false vacuum, the
seed bubble, in a region of true vacuum separated by a domain wall of
negligible thickness with surface energy density $\sigma$ as pictured
in Fig. \ref{fig:diag}. The thin wall approximation assumes that the
false vacuum is de Sitter space, the true vacuum is Schwarzschild, and
the stress energy tensor is discontinuous at the domain wall 


The classical solutions are discussed extensively in \cite{bgg} and it suffices to repeat a few key results. The mass $M$ of the bubble is the usual Schwarzschild parameter $m$ in the static foliation. This mass can be rewritten as

\begin{eqnarray}
M=\frac{\Lambda^2 r^3}{2G}+4\pi\sigma r^2\sqrt{1+\dot{r}^2-\Lambda^2 r^2}-8\pi^2 G \sigma^2 r^3
\end{eqnarray}
where $\Lambda$ is the cosmological constant, $G$ is Newton's constant, $\sigma$ is the surface energy density of the wall, and $r$ is the radial coordinate in the static de Sitter and Schwarzschild foliations. Note that the naive limit $r\rightarrow0$ appears to give zero mass solutions.

One can use the junction formalism developed in \cite{Berezin}. The basic strategy is to place a coordinate system on the wall and demand continuity of the metric tensor. Then utilizing Einstein's equation, the rescaled radial coordinate, $z$, obeys the following equation 

$$
\dot{z}^2+V(z)=E
$$
which is identical to that of a particle moving in a one dimensional potential. We know that if $V>E$, then two solutions exist, but classically the particle cannot move across the barrier. We can, however, have quantum tunneling between the two solutions.

\subsection{The two solutions}

As previously mentioned, two possible solutions exist for the
classical motion provided $M < M_{cr}$ where $M_{cr}\sim \Lambda \hat{r}^3$ is the characteristic mass of the problem \cite{bgg}. Here $\hat{r}$ is the radius of the bubble wall. Type (a) solutions are bounded
solutions that begin at $\hat{r}=0$, expand to some $\hat{r}=r_{max}$ before collapsing
back to zero. These solutions avoid a classical singularity, as discussed in FGG, because the trajectory on the Kruskal diagram crosses to the right of the origin and a closed ``anti-trapped'' surface no longer exists. This point is further elucidated in \cite{aj}.

Second, there are bounce solutions in which $\hat{r}$ approaches infinity in the asymptotic past, falls to some minimum value, and expands again to approach infinity in the asymptotic future. The Penrose theorem implies that this space-time must have emerged from an initial singularity, since the bubble radius grows beyond $(\Lambda)^-{\frac{1}{2}}$. The way to avoid this classical singularity yet still produce an inflating universe is to consider tunneling between the two solutions. The two solutions are of identical mass and thus identical energy. This is the FGG process.   

The tunneling probability can be calculated using a functional
integral \cite{fgg} or a canonical quantization \cite{fmp}. In either
case, the probability of tunneling between the two solutions remains
finite as the mass of the seed bubble, an input parameter, is taken to zero.

\section{Visiting the thick wall}
Consider a scalar field minimally coupled to gravity in a quartic potential, described by the following action
\begin{eqnarray}
S=\frac{1}{2}m_P^2 \int d^4 x \sqrt{-g}(R -\nabla_{a}\phi\nabla_{b}\phi g^{ab} - 2 V(\phi))
\end{eqnarray}
where 
$V(\phi)=\lambda \phi^4 - \gamma \phi^3 + \frac{m_{i}^2}{2}\phi^2$,
$m_{i}$ is the inflaton mass, and $m_P$ is the reduced Planck mass. A particular potential is shown in Fig. \ref{fig:potential}.

We work in a $+2$ metric signature, in reduced Planck units where $\hbar=1,c=1$ and $m_P=\sqrt{8\pi G}^{-1}$. One can then nondimensionalize the problem by rescaling the coordinates, for example by using $r^*=r m_P$. This rescales the potential to

$$
V(\phi)=\lambda \phi^4 - \gamma \phi^3 + \frac{m_{i}^2}{2 m_P^2}\phi^2
$$
Here $\phi,\lambda$ and $\gamma$ are all dimensionless. In what
follows, all coordinates and quantities are dimensionless with $m_i =
m_P$.  Keeping $m_i$ fixed allows us to explore the properties of the
bubble solutions without allowing the overall scale of the potential
to vanish (in that case one does expect solutions with $M$ approaching
zero to be possible). Fixing $m_i$ to the value $m_P$ is convenient
for the dynamic range of our numerical work is also a common choice in inflationary models. We use
standard spherical, $(t,r,\theta,\phi)$, coordinates. Under the
assumption of spherical symmetry, the spacetime line element takes the
form 
\begin{eqnarray}
ds^2 = -\alpha^2 (r,t) dt^2 + a^2(r,t) dr^2 + r^2 d\Omega^2
\end{eqnarray}
Note that we have not forced the metric in any region to take the de Sitter or Schwarzschild form, although we do require that the spacetime is asymptotically flat at large $r$. Here $r$ is both a coordinate and the measure of proper area. The stress-energy tensor of a scalar field in a potential is
\begin{eqnarray}
T_{ab}=\partial_{a}\phi\partial_{b}\phi-\frac{1}{2}g_{ab}\left(\partial_{c}\phi\partial_{d}\phi g^{cd}+2V(\phi)\right)
\end{eqnarray}

Defining mass in the thick-wall case is more involved, since we no
longer have a region of exact Schwarzschild space or a fixed wall
position where one can place an observer. Instead, we focus on the ADM
mass, which is defined at spatial infinity for asymptotically flat
spacetimes. This is the most relevant mass for tunneling calculations
\cite{albrechtsorbo}. This mass is defined as \cite{chop}  
\begin{eqnarray}
\label{meq}
M=2\pi \int_{0}^{\infty} dr r^2 \left[\left(\frac{\phi'}{a}\right)^2+\left(\frac{\dot{\phi}}{\alpha}\right)^2 + 2 V(\phi) \right]
\end{eqnarray}
where prime denotes differentiation with respect to $r$ while dot denotes differentiation with respect to time. 

The $00$ and $11$ components of Einstein's equation and the scalar
field equation are used in the simulation, while the $22$ equation is used as a consistency check.


We want to find solutions for a bubble of false vacuum embedded in
true vacuum, i.e. we want to find the radial profiles and time
evolution of $\phi,\alpha$ and $a$. This is done by demanding that
$\phi (r=r_{min})$ take the value of the false minimum of the
potential, so that $V(\phi_m)$ acts as a cosmological constant near
the origin. We also investigated cases in which the scalar field was
not initially at the minimum of the potential. This did not change our
conclusions regarding $M\rightarrow 0$, which are mainly sensitive to
the behavior of the bubble solutions as at larger values of $r$.
At large $r$, demanding true vacuum implies $\phi\rightarrow0$ since $V(\phi=0)=0$. 
  These conditions are labeled in Fig. \ref{fig:potential}.

\begin{figure}[b]
\includegraphics{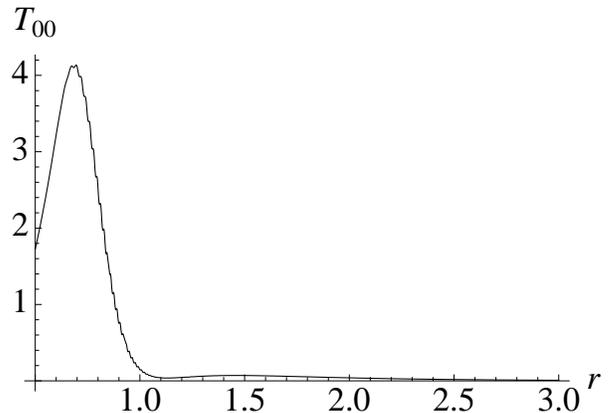}
\caption{\label{fig:t00} $T_{00}$, demonstrating the wall ``thickness'', at a given time.}
\end{figure}

\subsection{The turning point}
For the purposes of this article it suffices to examine the properties 
of the classical solutions relevant to the FGG
process. There is no need to find the tunneling solutions and the corresponding
tunneling actions to make our points. Furthermore, we can understand
the relevant properties of these solutions (namely the ADM mass)
simply by finding the solution at its turning point, which further
simplifies our calculations.   At the turning point $\dot{\alpha}=\dot{a}=\dot{\phi}=0$ but second order time derivatives are nonzero. In this case, equation \ref{meq} reduces to

\begin{equation}
M=2\pi\int_{0}^{\infty}  r^2 \left[\left(\frac{\phi'}{a}\right)^2+ 2
  V(\phi) \right]dr
\label{MatTurningPoint}
\end{equation}
Using Einstein's equation, this can be rewritten as (again setting $m_{i}=m_P$),
$$
M=\frac{4\pi}{3} \int_{0}^{\infty} \left(\frac{2ra'}{a^3}-\frac{1}{a^2}+1\right)dr
$$
so we see that $dm/dr=0$ only for the Schwarzschild metric, as
expected. Additionally, this formulation gives $\rho_{vac} V_{bubble}$
for the de Sitter contribution, where $\rho_{vac}$ is the energy
density of the vacuum, which is proportional to $\Lambda$. By
inspection of Eqn. \ref{MatTurningPoint} (which is positive definite since $V(\phi)$ is everywhere positive)
one can see that we do not expect to find
$M\equiv0$ solutions, but there is no apparent reason why a smooth
limit to zero should not exist.

There is freedom to specify the spatial profile of $\ddot{\phi}$ at
the turning point, which we choose to be $\ddot{\phi}=c/r^2$ for a
constant $c$. This is consistent with spherical symmetry and is
sufficiently localized to maintain an asymptotically flat spacetime.
Choosing such an ansatz simply enforces locality of the bubble and does not affect the generality of our
conclusions. 

\subsection{The two solutions}
\begin{figure}[b]
\includegraphics{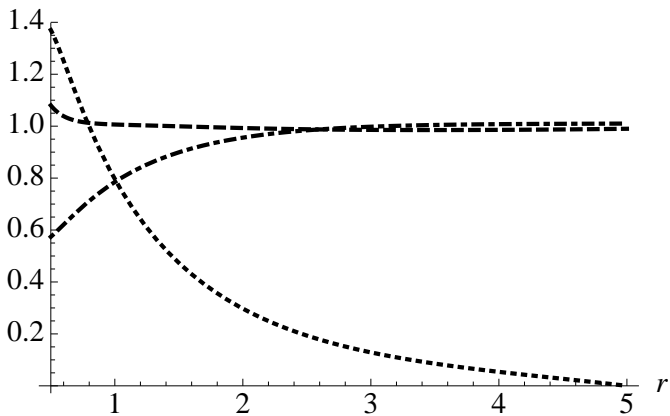}
\caption{\label{fig:exp} Turning point slice of geometry and field for
  the expanding solution. Here $\phi$ is dotted, $a$ is dash-dotted,
  and $\alpha$ is dashed. }
\end{figure}

\begin{figure}[b]
\includegraphics{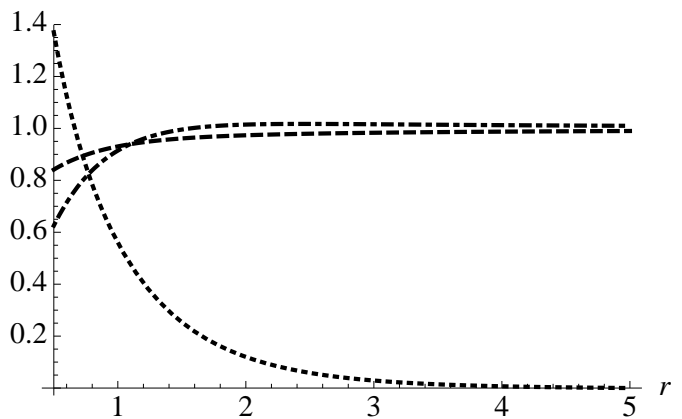}
\caption{\label{fig:coll} Turning point slice of the collapsing solution. Here $\phi$ is dotted, $a$ is dash-dotted, and $\alpha$ is dashed.}
\end{figure}

Evolving forward in time from the turning point solution is used to classify the solution character. The energy density of the expanding solution expands into the domain as the metric functions approach de Sitter (see \ref{fig:t00}). The energy density of the collapsing solution collapses immediately toward the origin while the metric functions approach pure Schwarzschild. A plot of $T_{00}(t=0)$, showing the ``thickness'' of the wall, for the expanding solution is given in figure \ref{fig:t00}. Plots of the field and metric functions for expanding and collapsing solutions are given in figures \ref{fig:exp} and \ref{fig:coll}, respectively.

\subsection{Results of trying to take $M\rightarrow0$}
Conceptually, there are two ways in which this can be done. From a
cosmological perspective, we can fix the inflaton potential and
attempt to take the mass to zero by changing the initial condition on
$\ddot{\phi}$.  

\begin{figure}[b]
\includegraphics{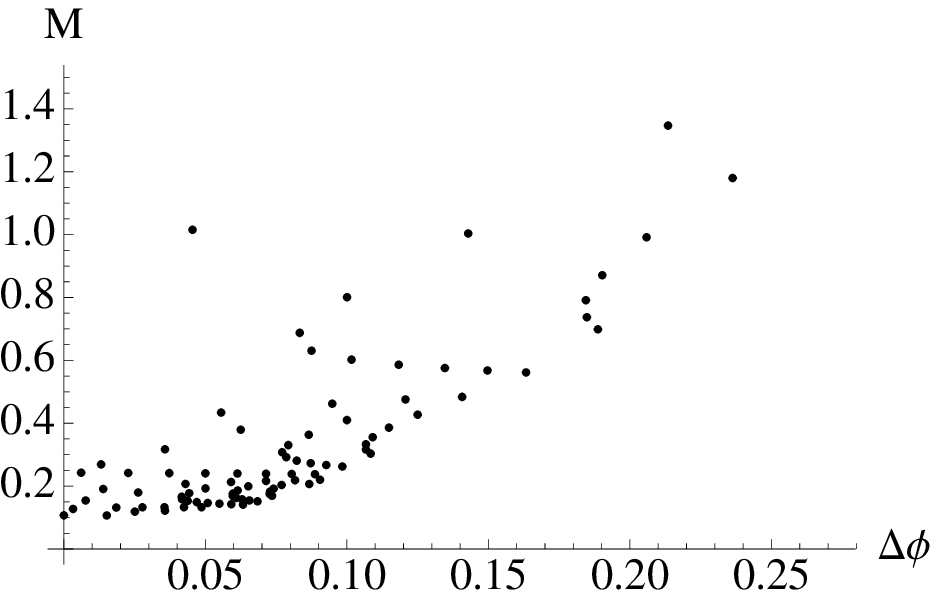}
\caption{\label{fig:mdepdP} Mass of the collapsing solution, evaluated
  at the turning point, as a function of $\Delta\phi$. Points represent
a scan of potential parameters $\lambda$ and $\gamma$. Small values of
$M$ were not found in the scan.} 

\includegraphics{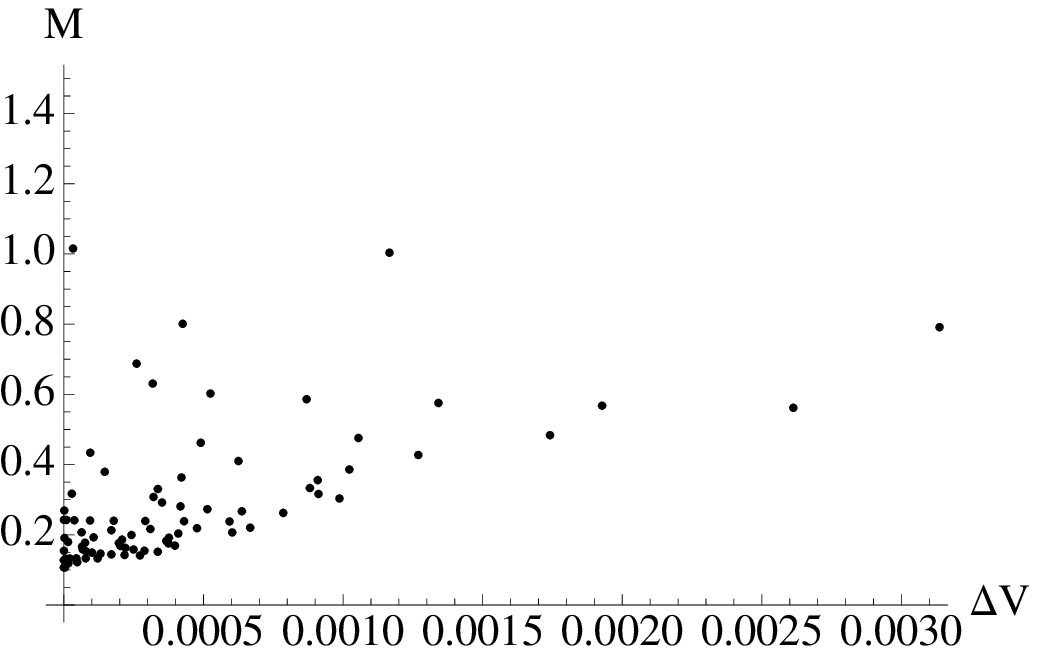}
\caption{\label{fig:mdepdV} Mass of the collapsing solution, evaluated
  at the turning point, as a function of $\Delta V$. Points represent
a scan of potential parameters $\lambda$ and $\gamma$. Small values of
$M$ were not found in the scan.}

\end{figure}

On the other hand, we can tune the two constants in the potential
while keeping a fixed $\ddot{\phi}$. (As discussed and motivated above, we are
keeping the overall scale $m_i$ fixed for this investigation.) We begin with a parameter scan
over three orders of magnitude, i.e. ranging the values of $\lambda$
and $\gamma$ from $0.1\rightarrow10$. Let the value of the field at
the false minimum and the maximum be $\phi_{min}$ and $\phi_{max}$,
and the potential evaluated at these points be $V_{min}$ and
$V_{max}$, respectively. Figures \ref{fig:mdepdP} and \ref{fig:mdepdV}
show how the mass of the collapsing turning point solution depends on
$\Delta\phi=\phi_{min}-\phi_{max}$ and $\Delta V=V_{max}-V_{min}$.

\section{Discussion} 
Inspecting Eqn. \ref{meq}, we see that the integrand, $dm/dr$, will
not be zero unless $\phi$ is constant and $V(\phi)=0$, i.e. exactly
Schwarzschild space. The crux of our argument is that real potentials
and fields do not admit nicely separated solutions; the field spills
over into the whole domain and affects the metric functions,
preventing the $M\rightarrow0$ limit that appears to exist in the
thin-wall formalism.  

We still attempted to push the mass smoothly to zero. However, as
Figures \ref{fig:mdepdP} and  \ref{fig:mdepdV} show, we are unable to
push the mass below about $0.1$. This is with a fixed overall scale
set by choosing $m_i=m_P$. The point is not the value of $m_i$
(setting $(m_i/m_P)^2=10^{-3}$ does not affect our conclusions), but
that we have fixed an overall scale.  

In any spherically symmetric problem, there is the issue of what
happens at $r=0$. While the numerics cannot evolve such a point, we
can make progress analytically by assuming we approach exact de Sitter
space, in which the metric functions are regular at the origin. The
scalar field potential can then be expanding about the minimum to
second order in $\phi$. The problem is then analytically tractable and
solutions give positive mass contributions. Thus, our calculation of
$M$ really is a lower bound.  

Lastly, there is the issue of horizons in the computational domain. If
we had exact de Sitter, there would be a de Sitter horizon at
$r=(\Lambda)^{-\frac{1}{2}}$. We explicitly chose $\Lambda$, via
parameters in the potential, to be small so that the horizon was
significantly outside of the computational domain. The Schwarzschild
coordinate horizon is sufficiently inside $r_{min}$ for the attempted
$M\rightarrow 0$ calculations.  

\section{Conclusion}
We considered classical solutions relevant to the Farhi-Guth-Guven
tunneling process. For a generic quartic potential we are unable to
take the mass of our turning point solutions smoothly to zero. 
Other authors have shown using the thin wall approximation that the FGG tunneling
amplitude remains finite as $M\rightarrow0$, and this strange behavior
has been used to question the physicality of the FGG process.  The
absence of $M\rightarrow0$ solutions in our more realistic thick wall
calculations suggest that the $M\rightarrow0$ behavior is an artifact
of thin wall approximation and should not be used to argue that the FGG
process is unphysical. 

\section{Acknowledgements}
This work was supported in part by a McCormick Fellowship (AU), and
DOE grant DE-FG03-91ER4067 (AA and AU). AA thanks the University of
Chicago department of Astronomy and Astrophysics and KICP for
hospitality during his sabbatical, when most of this work was completed. We thank Bob Wald and Emil Martinec for fruitful discussions. We also thank A. Aguirre and M. Johnson for helpful conversations, and the Perimeter Institute, where many of these conversations took place.

\bibliography{twref}

\end{document}